\documentclass[aps,prx,reprint,superscriptaddress,longbibliography,floatfix,amsmath,amssymb]{revtex4-2}

\usepackage[utf8]{inputenc}
\usepackage[T1]{fontenc}
\usepackage{graphicx}
\usepackage{dcolumn}
\usepackage{bm}
\usepackage{mathtools}
\usepackage{multirow}
\usepackage{makecell}
\usepackage{physics}
\usepackage{upgreek}
\usepackage[colorlinks=true,linkcolor=blue,urlcolor=blue,citecolor=blue]{hyperref}

\begin{document}

\title{Large superconducting diode effect indicates nonuniform triplet pairing in UTe$_2$}

\author{Zheyu~Wu}
\thanks{These authors contributed equally to this work.}
\affiliation{Cavendish Laboratory, University of Cambridge, JJ Thomson Avenue, Cambridge, CB3 0US, UK}
\affiliation{National High Magnetic Field Laboratory, Florida State University, Tallahassee, Florida 32310, USA}
\author{Daniel~Shaffer}
\thanks{These authors contributed equally to this work.}
\affiliation{Department of Physics, University of Wisconsin-Madison, Madison, Wisconsin 53706, USA}
\author{Dmitry~V.~Chichinadze}
\thanks{These authors contributed equally to this work.}
\affiliation{Department of Physics, Washington University in St. Louis, St. Louis, MO, 63130 USA}
\author{Mengmeng~Long}
\affiliation{Cavendish Laboratory, University of Cambridge, JJ Thomson Avenue, Cambridge, CB3 0US, UK}
\author{Hanyi~Chen}
\affiliation{Cavendish Laboratory, University of Cambridge, JJ Thomson Avenue, Cambridge, CB3 0US, UK}
\author{Andrej~Cabala}
\affiliation{Charles University, Faculty of Mathematics and Physics, Department of Condensed Matter Physics, Ke Karlovu 5, Prague 2, 121 16, Czech Republic}
\author{Theodore~I.~Weinberger}
\affiliation{Cavendish Laboratory, University of Cambridge, JJ Thomson Avenue, Cambridge, CB3 0US, UK}
\author{Maurice~Bal}
\affiliation{High Field Magnet Laboratory (HFML-EMFL), Radboud University, Toernooiveld 7, 6525 ED Nijmegen, The Netherlands}
\author{Steffen~Wiedmann}
\affiliation{High Field Magnet Laboratory (HFML-EMFL), Radboud University, Toernooiveld 7, 6525 ED Nijmegen, The Netherlands}
\author{Dave~Graf}
\affiliation{National High Magnetic Field Laboratory, Florida State University, Tallahassee, Florida 32310, USA}
\author{Vladimir~Sechovsk\'y}
\affiliation{Charles University, Faculty of Mathematics and Physics, Department of Condensed Matter Physics, Ke Karlovu 5, Prague 2, 121 16, Czech Republic}
\author{Michal~Vali{\v{s}}ka}
\affiliation{Charles University, Faculty of Mathematics and Physics, Department of Condensed Matter Physics, Ke Karlovu 5, Prague 2, 121 16, Czech Republic}
\author{F.~Malte~Grosche}
\affiliation{Cavendish Laboratory, University of Cambridge, JJ Thomson Avenue, Cambridge, CB3 0US, UK}
\author{Alex~Levchenko}
\affiliation{Department of Physics, University of Wisconsin-Madison, Madison, Wisconsin 53706, USA}
\author{Alexander~G.~Eaton}
\email[Contact author: ]{alex.eaton@phy.cam.ac.uk}
\affiliation{Cavendish Laboratory, University of Cambridge, JJ Thomson Avenue, Cambridge, CB3 0US, UK}

\date{\today}

\begin{abstract}
The superconducting diode effect (SDE) is characterized by unequal critical current magnitudes for opposite-polarity current flow, owing to broken spatial-inversion and time-reversal symmetries. For a superconducting pair wavefunction to spontaneously break inversion symmetry implies the formation of Cooper pairs with nonzero momentum, yielding a nonuniform superconducting state. For singlet superconductors, the formation of such a (FFLO) state has long been understood to provide an escape from high field Pauli pair-breaking -- how and why such an effect might occur for equal spin pairing is an open question. Here we report the observation of a large SDE in bulk specimens of the three-dimensional heavy-fermion triplet superconductor $\text{UTe}_2$, with a peak diode efficiency of 33\% at the high magnetic field strength of 33.6~T, representing the highest-field observation of the SDE to date. Crucially, this nonreciprocal transport is strictly confined to a narrow field regime where superconductivity coexists with a field-induced spin-polarized state near a first-order metamagnetic phase boundary. The absence of the SDE away from this slender phase-intersection argues strongly against purely extrinsic vortex pinning asymmetries. Instead, our findings indicate spontaneous inversion symmetry breaking accompanied by finite-momentum Cooper pairing. We model this behavior using a phenomenological Ginzburg-Landau theory, demonstrating that the intertwining of magnetic and triplet superconducting orders can stabilize an odd-parity FFLO state.
\end{abstract}

\maketitle

\section{\label{sec:intro}Introduction}

Since the prediction of nonuniform superconducting states in 1964 by Fulde, Ferrell, Larkin and Ovchinnikov \cite{FF, LO} -- broadly referred to as FFLO states -- evidence for their existence has been found in multiple candidate materials. These include organic superconductors (SCs) \cite{Wosnitza18, Agosta18}, cuprates \cite{Agterberg20}, noncentrosymmetric SCs with helical states \cite{SmidmanAgterberg17}, SCs coexisting with magnetic order including heavy fermion SCs such as CeCoIn\(_5\) \cite{AgterbergSigrist09, YanaseSigrist09, Aperis10, Kenzelmann17}, as well as in cold atomic systems \cite{Radzihovsky10, KinnunenTorma18}. 
With more recent imaging techniques, e.g., scanning Josephson tunneling microscopy (SJTM), pair density waves (PDWs) -- nonuniform SCs that either do not break time reversal symmetry or break it spontaneously \cite{Agterberg20}, as opposed to FFLO states that occur with explicit time reversal symmetry breaking (TRSB) -- have been reported on surfaces of multiple SCs \cite{Wang26}, including cuprates~\cite{hamidian2016detection}, kagome SCs \cite{DengThomale24}, and heavy-fermion UTe\(_2\) \cite{GuDavis23, ZhuEatonMadhavan26}. Related nonuniform superconductivity with intra-unit-cell modulation that does not break translation symmetries has also been observed in Fe-chalcogenides \cite{KongPapaj25, Wei25}.

Despite the multitude of indirect experimental observations supporting the existence of FFLO states, conclusive identification of bulk nonuniform superconductivity -- characterized by finite-momentum Cooper pairs -- remains challenging due to the difficulty of resolving the bulk modulation of the SC order parameter. While recent advances in imaging techniques have partly addressed this issue, such measurements are constrained to probe only the material surface, making it challenging to rule out surface reconstruction effects as reported for example in the case of UTe\(_2\) \cite{KengleRosa24, TheussPaglione24, Li26, TalaveraAokiYanase26}.
A promising alternative bulk transport probe for identifying FFLO states with inversion symmetry breaking (ISB), the so-called FF-type FFLO and PDW states \cite{Agterberg20}, is provided by the superconducting diode effect (SDE) \cite{Nadeem23, NagaosaYanase24, Ma25, shaffer2025theoriessuperconductingdiodeeffects}. SDE, defined by unequal critical currents \(J_c^+\) and \(J_c^-\) for opposite current flow directions, initially attracted much attention precisely as a possible signature of a helical FFLO state in a noncentrosymmetric heterostructure \cite{AndoYanase20}; related effects in Josephson junctions have also been proposed \cite{YangAgterberg00, KaurAgterbergSigrist05, ZhuangShafferHasanLevchenko25} and used \cite{PalDavydovaFuParkin22} as methods of characterizing FFLO states. However, it has since been pointed out that extrinsic vortex diode effects (VDE), another form of SDE known but unnamed for a long time \cite{EdwardsNewhouse62, SwartzHart67}, can account for the critical current nonreciprocity without invoking finite-momentum Cooper pairs \cite{GutfreundBuzdin23, SundareshVayrynen23, MollGeshkenbein23, HouMoodera23}. Ruling out such extrinsic effects is, therefore, crucial for robustly identifying an FFLO state.

Here we report SDE in a bulk UTe\(_2\) sample, with the diode efficiency $\eta = (J_c^+ - |J_c^-|)/(J_c^+ + |J_c^-|)$ [the definition shown in Fig.~\ref{fig:intro}(e)] reaching a magnitude $|\eta| = 33\%$ near a field-induced first order metamagnetic phase transition into a polarized paramagnetic (PPM) state at $\mu_0 H\approx 33$~T, making it the highest-field observation of SDE to date. This is also a rare example of SDE in a bulk 3D sample, besides e.g. Refs.~\cite{SwartzHart67, Moll25}, most other reports being in 2D or thin-film samples.
Most significantly, the absence of SDE away from a narrow region close to the metamagnetic transition strongly disfavors a purely extrinsic vortex effect due to either bulk or surface pinning asymmetry, instead implying spontaneous ISB with accompanying nonuniform superconductivity. Moreover, we establish the coexistence of superconductivity within the PPM phase over a narrow field region in which the SDE is observed, suggesting the mixing of superconducting and magnetic orders as a likely source of the ISB. This ISB may be either due to some domain structure and/or an intrinsic FFLO state due to the coexistence of the two orders, which we model phenomenologically using a Ginzburg-Landau (GL) theory below.

In addition to other signatures of nonuniform SC in UTe\(_2\), in both surface SJTM/STM experiments in the low field (SC1) phase \cite{GuDavis23, ZhuEatonMadhavan26} 
and anomalous vortex transport in a mixed region between SC1 and the field-induced SC2 phase \cite{Aoki_Hard, Tokiwa-vortexPRB23, ZhangMoll25, Tony26memory}, the observation of SDE provides the strongest evidence yet for a bulk (FF-like) nonuniform SC state in UTe\(_2\).
Given the abundant evidence for triplet pairing in UTe\(_2\) \cite{YoonPaglione24, KitagawaAoki24, LiMaple25, matsumura2023NMR-aoki, Aoki_UTe2review2022, lewin2023review}, this is a rare candidate for a triplet FFLO state besides CeCoIn\(_5\) \cite{AgterbergSigrist09, YanaseSigrist09, Aperis10}, adding to the long list of possible unconventional phases in this material \cite{Jiao2020chiral, Aoki_UTe2review2022, lewin2023review}. 

\section{Superconducting diode effect at the overlap of superconductivity and magnetism}\label{sec:exp}

\begin{figure*}[t]
\centering
\includegraphics[width=\textwidth]{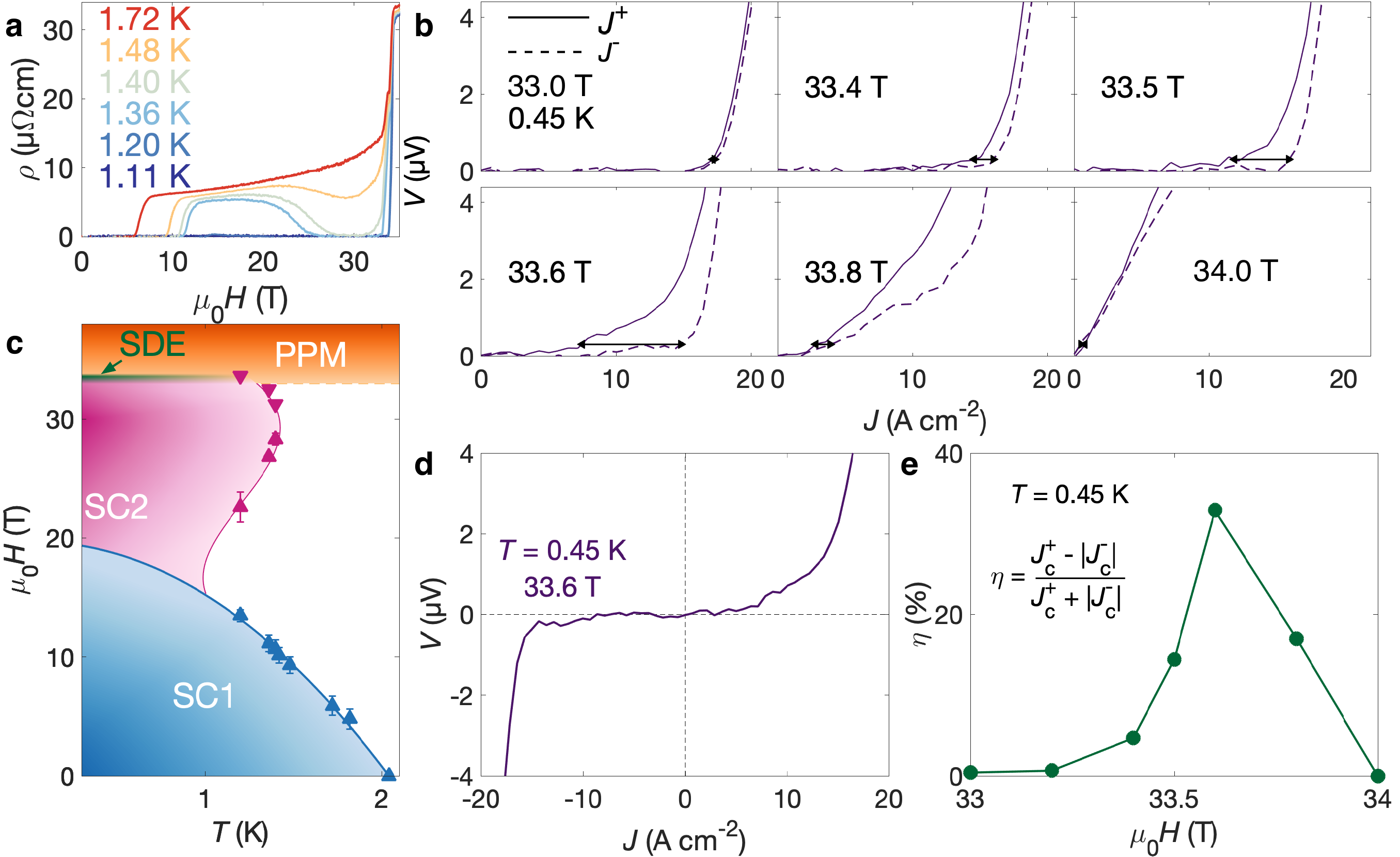}
\caption{\textbf{Superconducting diode effect in UTe$_2$.} (a) Resistivity $\rho$ as a function of applied magnetic field \textbf{H}~$\parallel$~[010] at various temperatures $T$. Throughout this article, all currents were sourced along [100]. At low $T$ re-entrant SC2 superconductivity occurs around 20-33~T, above which a sharp increase in $\rho$ marks a metamagnetic transition into a polarized paramagnetic (PPM) state. (b) Voltage $V$ as a function of dc current density $J$ for positive (solid curves) and negative (dashed curves) polarity current sweeps, at indicated steady $H$ values. At 33~T, $V(J^+) = V(J^-)$ within resolution. By contrast, between 33.4-33.8~T this is no longer the case, with very different forms of $V(J^+)$ compared to $V(J^-)$. (c) Low temperature UTe$_2$ phase diagram for \textbf{H}~$\parallel$~[010]~\cite{tony2024enhanced}. All points are from this study, with error bars calculated as the distance between the transition and the nearest data above the noise level. Green coloring indicates the small region where SC2 overlaps with the PPM state (see Fig.~\ref{fig:Tdep}), in which we observe the SDE. (d) Full voltage profile for opposite polarity currents at 33.6~T, where it is clear that the magnitude of the critical current $J_c^- > J_c^+$. (e) SDE efficiency $\eta$ (defined inset) as a function of $H$. A maximal $\eta$ of 33\% is observed at 33.6 T.}
\label{fig:intro}
\end{figure*}

We plot the electrical resistivity $\rho$ of UTe$_2$ versus applied magnetic field strength $\mu_0H$ in Fig.~\ref{fig:intro}(a) at various temperatures, with \textbf{H} aligned along the hard magnetic $b$-axis and \textbf{J} along the $a$-axis. At low temperatures $T \lessapprox 1.1$~K, $\rho(H) = 0$ up until around 34~T, where a steep upturn occurs at a metamagnetic phase transition as the material enters the high-spin PPM state, characterized by an abrupt increase in the bulk magnetization of 0.5 Bohr magnetons per uranium site~\cite{Miyake2019, NiuPhysRevResearch.2.033179, Valiska24, Tony25meta}. At higher temperatures re-entrant superconductivity is observed, distinguishing two separate superconducting phases (numbered SC1 and SC2)~\cite{lewin2023review}. In Fig.~\ref{fig:intro}(b) we compare the $J-V$ characteristics of the SC2 state at various fixed $H$ values for positive- and negative-polarity $J$-sweeps. At 33~T we observe $V(J^+) = V(J^-)$, within experimental resolution. By contrast, a large nonreciprocity in $\pm J_c$ develops upon increasing $H$, yielding a sizeable SDE. This is maximal at $\mu_0H = 33.6$~T, exhibiting an SDE efficiency of 33\%. Then, by 34~T, SC2 is fully suppressed and Ohmic, polarity-symmetric magnetoconductivity is observed. Experimental methods, the evolution of the SDE with field angle, and its reproduction in a second sample (S2) are presented in Appendix~\ref{app:methods}.

\begin{figure*}[t]
\centering
\includegraphics[width=\textwidth]{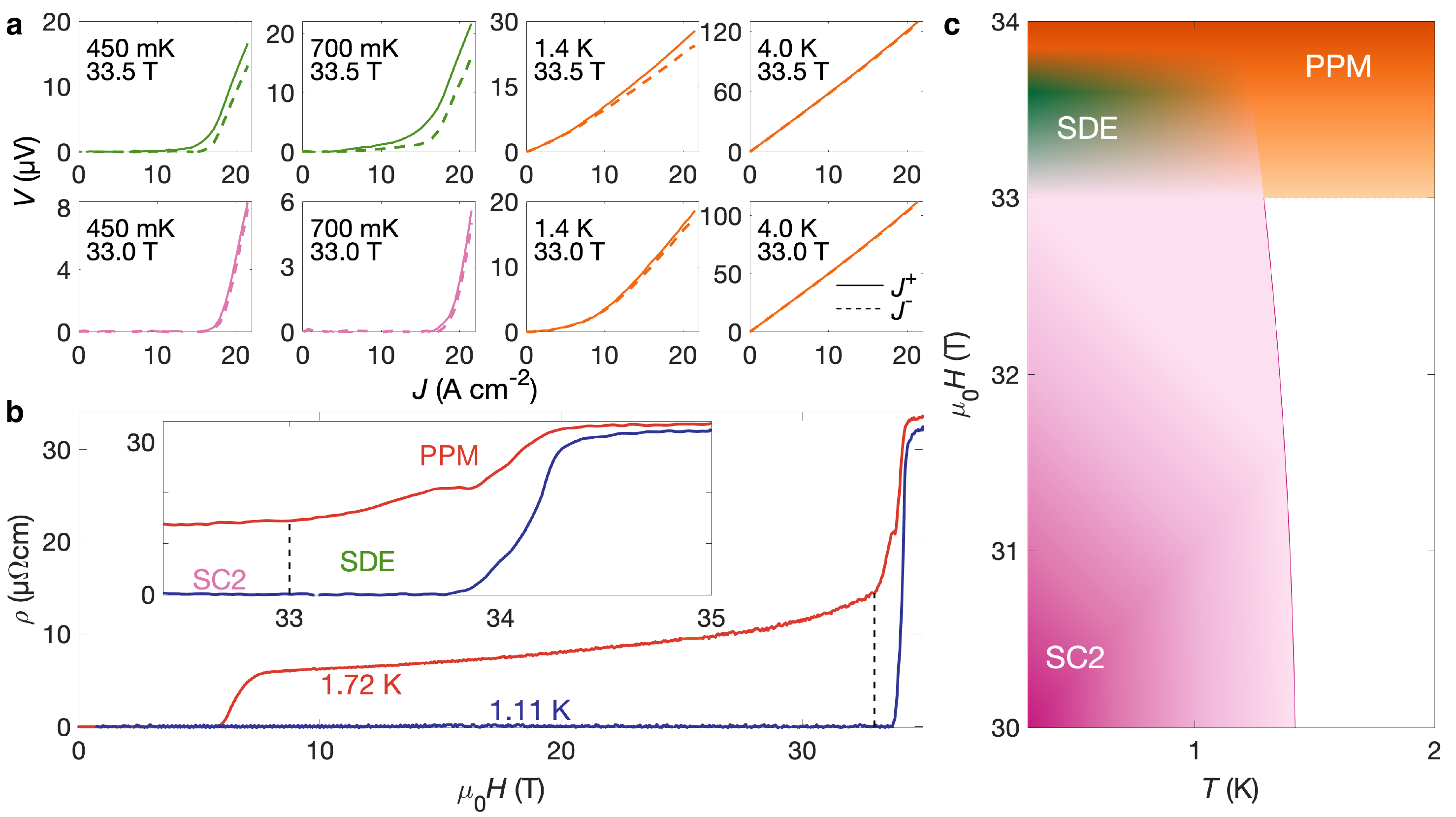}
\caption{\textbf{Coexistence of superconductivity and magnetism.} (a) Positive (solid) and negative (dashed) polarity current sweeps at various temperatures $T$ in fixed fields of 33.5~T (top row) and 33.0~T (bottom row). SDE is visible at low $T$ at 33.5~T. At 4.0~K (above $T_c$) Ohmic, reciprocal $V(J)$ is observed. (b) The 1.11~K and 1.72~K resistivity $\rho$ data replotted from Fig.~\ref{fig:intro}(a), with zoomed inset. The small field region over which the SDE is observed between 33-34~T shows $\rho = 0$ at 1.11~K. By contrast, after warming above the $T_c$ of SC2, at 1.72~K the metamagnetic transition -- characterized by the sharp increase in $\rho$ (marked by dashed line) -- is observed to start at 33.0~T. This indicates that the SDE occurs when SC2 and PPM share a region of overlap. (c) Schematic phase diagram of UTe$_2$ at low $T$ for \textbf{H} $\parallel b$. The SDE (colored green) only occurs in a narrow portion of the phase landscape in which the SC2 and PPM phases overlap.}
\label{fig:Tdep}
\end{figure*}

In Fig.~\ref{fig:Tdep}(b) we replot the 1.11~K and 1.72~K $\rho(H)$ curves from Fig.~\ref{fig:intro}(a), to evaluate how the manifestation of a large SDE for $\mu_0 H > 33$~T relates to the metamagnetic phase boundary. At 1.72~K the sample transitions from SC1 to the normal state at 5.8~T, and then does not enter SC2 at any higher $H$. At 33.0~T, $\rho$ bends up sharply (marked by dashed line), indicating that the metamagnetic transition has been crossed into the PPM state. However, interestingly, upon taking a closer look in the zoomed inset it is clear that at 1.11~K $\rho$ remains zero up until 33.8~T. This implies that there is a small region of overlap between the SC2 and PPM states from 33.0 to 33.8~T -- corresponding to the field domain over which the SDE occurs (Fig.~\ref{fig:intro}). This strongly suggests that the interplay of superconductivity and magnetism drives the manifestation of the SDE in UTe$_2$.

\begin{figure*}[t]
\centering
\includegraphics[width=0.85\textwidth]{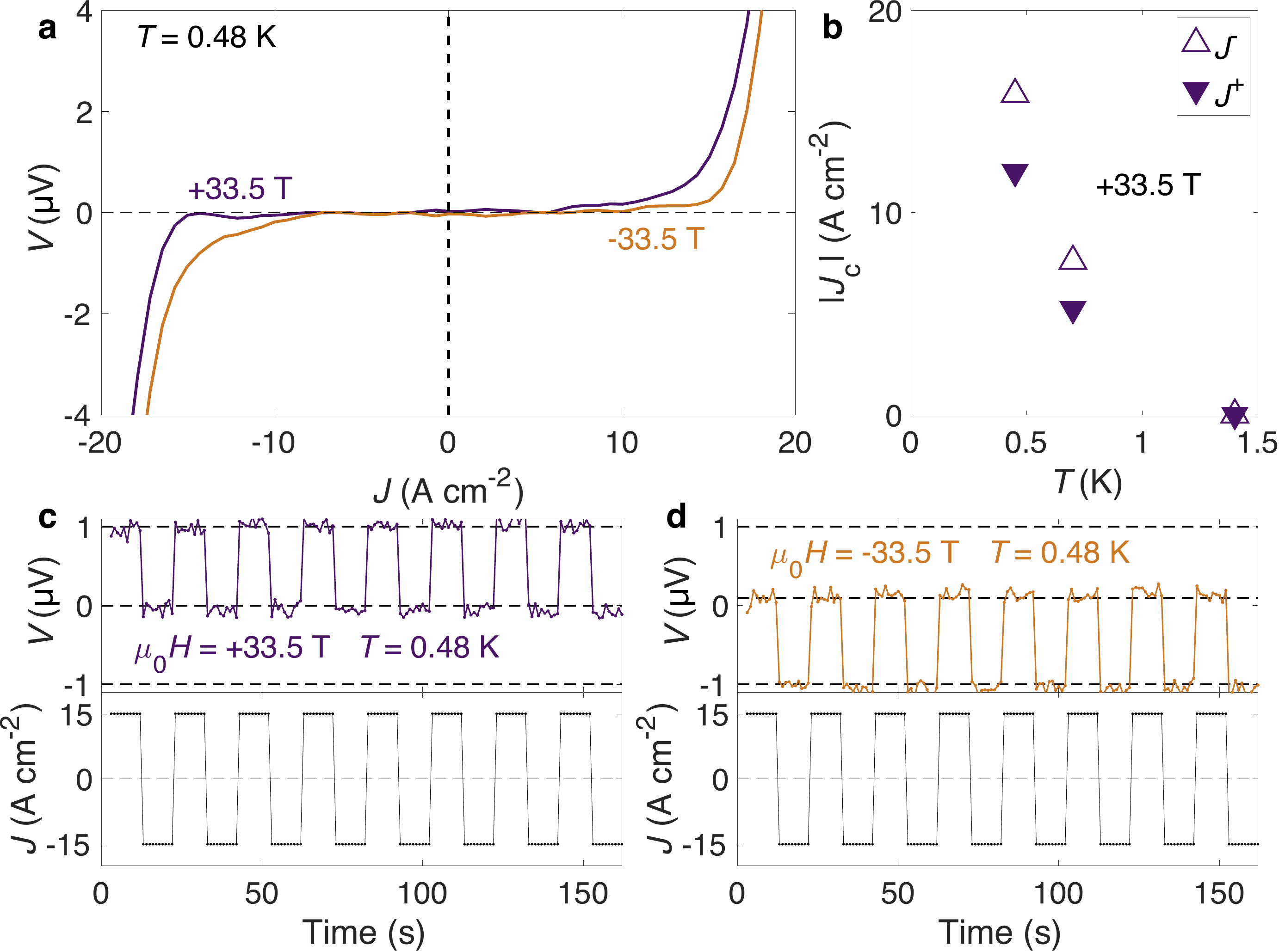}
\caption{\textbf{Supercurrent rectification.} (a) $V(J)$ measured for positive (purple) and negative (ochre) polarity magnetic field orientations of $\abs{\mu_0H} = 33.5$~T. The two curves are field-antisymmetric -- that is, one can be mapped onto the other by reversing the signs of both $J$ and $V$, $V_{-H}(J)=-V_{+H}(-J)$. (b) Magnitudes of the critical current densities $J_c^+$ (filled triangles) and $J_c^-$ (open triangles) as a function of temperature at $\mu_0H = +33.5$~T, extracted from the $V(J)$ sweeps in Fig.~\ref{fig:Tdep}(a). $\abs{J_c^-}$ exceeds $J_c^+$ at all temperatures where the critical currents are finite, and both are suppressed to zero by 1.4~K. (c),(d) Supercurrent rectification for $\mu_0H = +33.5$~T (c) and at $\mu_0H = -33.5$~T (d), as $J$ is alternated between $\pm15$~A\,cm$^{-2}$. By sitting at this point in the $V(J)$ curve, the potential difference across the sample can be switched between 0 and $\pm1 \upmu$V, depending on field polarity.}
\label{fig:rect}
\end{figure*}

In Fig.~\ref{fig:rect} we examine the profile of $V(J)$ upon switching the polarity of $H$. In Fig.~\ref{fig:rect}(a), it is clear that for  positive polarity magnetic field (purple curve), $J_c^- > J_c^+$. By contrast, for negative polarity field (ochre curve), the opposite is true with $J_c^- < J_c^+$. In panels (c) and (d) we bias the sample right at the edge of the nonreciprocal $J-V$ curves: at $\pm$33.5~T, switching $J$ between $\pm$15~A\,cm$^{-2}$ alternates $V$ between 0 and $\pm$1~$\upmu$V.



\section{Phenomenological Model of Spontaneous Intrinsic SDE}\label{sec:GL}

The experimental data shows that the SDE occurs in a narrow region of the phase diagram in which the SC2 and PPM states coexist, indicating spontaneous inversion symmetry breaking in the mixed state. To model this phenomenologically, we use an extension of the Ginzburg-Landau (GL) theory considered in Ref.~\cite{tony2024enhanced}. Taking \(M\) to be the magnetization field and \(\Delta\) to be the (possibly multicomponent) SC2 order parameter, the free energy \(\mathcal{F}[M,\Delta]=\mathcal{F}_M[M]+\mathcal{F}_{\Delta}[\Delta]+\mathcal{F}_{mix}[M,\Delta]\) has three terms:
\begin{align}\label{F}
    \mathcal{F}_M&=\frac{1}{2}\chi^{-1}M^2+\frac{1}{4}\beta M^4+\frac{1}{6}\delta M^6-MH\nonumber\\
    \mathcal{F}_{\Delta}&=(a_0(T)+a_H(H)+a_2q^2+a_4q^4)|\Delta|^2+b|\Delta|^4\nonumber\\
    \mathcal{F}_{mix}&=\frac{1}{2}\chi^{-1}_\Delta M^2|\Delta|^2+\gamma(q)M|\Delta|^2
\end{align}
Here \(\mathcal{F}_M\) and \(\mathcal{F}_{\Delta}\) model the pure magnetic and superconducting orders, with parameters chosen to approximately fit the metamagnetic phase transition and the critical temperature of SC2, respectively (see Appendix~\ref{app:GL}). 
Here \(q\) is the Cooper pair momentum (the corresponding GL terms originate from Fourier transforms of gradients of \(\Delta\)).
In the case of a multicomponent order parameter considered in the following section, the components can be parametrized as \(\Delta_m=|\Delta|d_m\), in which case the coefficients in front of \(|\Delta|^2\) terms in Eq. \eqref{F} should be understood as including cross terms of the form \(\Delta_m^*\Delta_n\), etc.

\begin{figure}[t]
\begin{centering}
\includegraphics[width=\columnwidth]{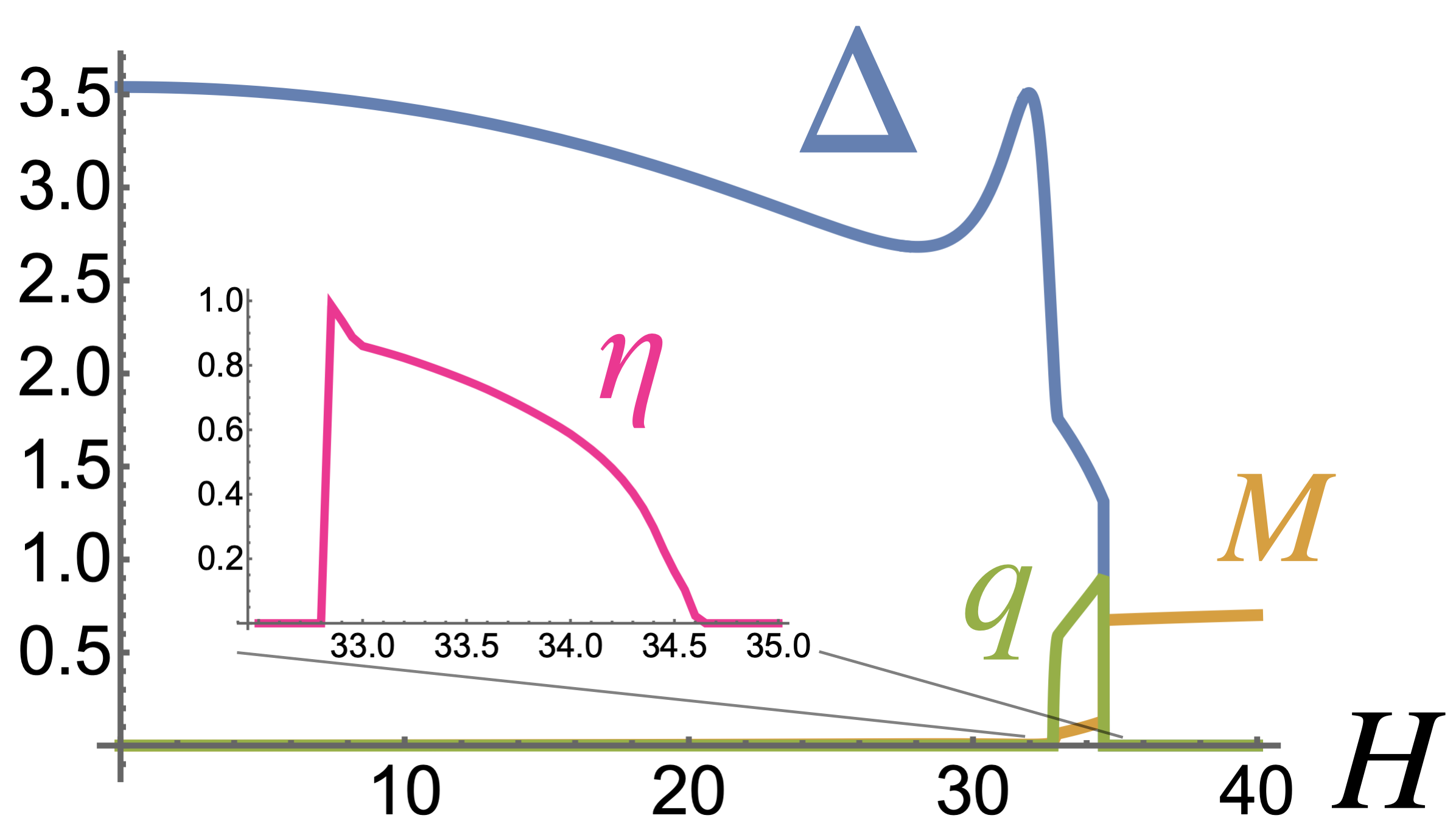}
\caption{\textbf{Ginzburg-Landau modeling of a triplet FFLO state.} Order parameters \(\Delta\), \(M\), and \(q\) versus applied magnetic field \(H\) (GL parameters listed in Appendix~\ref{app:GL}). A non-zero \(q\), indicating spontaneous ISB, occurs near the metamagnetic phase transition, accompanied by a non-zero diode efficiency \(\eta\), shown in the inset.}
\label{Fig:order}
\end{centering}
\end{figure}

The two orders in general interact via the mixed terms in \(\mathcal{F}_{mix}\). This includes the standard diamagnetic response of the superconductor through the \(\chi^{-1}_\Delta\gg0\) term, which strongly suppresses any coexistence. The crucial observation, however, is that an additional mixing term \(\gamma(q)=\gamma_2q^2\) is allowed by symmetry depending on the form of the order parameters, and may favor coexistence, provided \(q\neq 0\) in the ground state and \(\gamma_2\) is sufficiently negative. Such a state corresponds to \(\Delta\sim e^{iqr}\), i.e., an FF-type FFLO state, which breaks inversion symmetry.
We note that on symmetry grounds, ISB implies finite-momentum pairing in absence of fine-tuning; typically translation symmetry is also broken, unless \(\mathbf{q}\) is an element of the reciprocal lattice, in which case a so-called pair density modulation (PDM) state is realized \cite{KongPapaj25, Wei25}.
We comment on the possible microscopic origin of \(\gamma\) in the following section.
Figure \ref{Fig:order} shows the numerically computed ground state values of the order parameters \(\Delta\) and \(M\), and \(q\); the inset shows the diode efficiency \(\eta\); see Appendix~\ref{app:GL} for parameter values.
The metamagnetic phase transition occurs at \(\mu_0H=33\)~T in the normal state, but a phase characterized by non-zero \(q\) emerges just below this field in the SC state. Here we look at a fixed temperature \(T\), but \(T_c\) as a function of \(H\) is strongly correlated with \(\Delta(H)\).

\section{Possible Origins of SDE}\label{sec:origins}

Having presented the data and GL model for the SDE, we now discuss its possible origin and implications for the symmetry of the order parameter. 
We note that free energies similar to the one considered above have been considered in the context of uniform SC (i.e., \(q=0\)) in studies of TRSB in SC1 \cite{Nevidomskyy20, Hayes2021, WeiAgterbergSchmalian22}, as well as phenomenological studies of \(H_{c2}\) in SC2 and SC3 with non-unitary pairing \cite{Machida21, Machida23, lewin2025halo}. 
As noted in those works, at zero applied field a coupling of the form \(\gamma(q) M|\Delta|^2\) generally requires an SC order parameter that breaks TRS assuming \(\gamma(q)\) is an even function of \(q\). Since the UTe\(_2\) point group, \(D_{2h}\), only contains one dimensional irreducible representations (irreps), this generally requires SC2 to be a multicomponent phase with at least two SC components \(\Delta_n,\Delta_m\propto\Delta\) belonging to two distinct irreps with an imaginary phase between them (e.g., \(B_{1u}+i B_{3u}\) pairing, \(B_{2u}+i B_{3u}\) pairing, etc.).
For a single-band triplet superconductor, this generally implies non-unitary pairing with a gap function \(\Delta=i\sigma^y\boldsymbol{\sigma}\cdot\mathbf{d}\) with \(\mathbf{d}\)-vector \(\mathbf{d}\neq \mathbf{d}^*\), furthermore implying Cooper pairs possessing a non-vanishing Cooper pair spin \(\mathbf{s}=i\mathbf{d}\times\mathbf{d}^*\neq 0\) \cite{lewin2025halo}. 
However, we note that in a multicomponent SC, which is plausible in UTe\(_2\), non-unitary pairing is not necessary to produce a non-zero \(\gamma\) \cite{Ramires22}.

There are several possible microscopic mechanisms for generating a coupling of the form \(\gamma M|\Delta|^2\), the simplest being an exchange-like interaction \(\gamma\mathbf{M}\cdot\mathbf{s}=i\gamma\mathbf{M}\cdot\mathbf{d}\times\mathbf{d}^*\sim \gamma M\Delta^*_n\Delta_m\). Noting that UTe\(_2\) is a locally noncentrosymmetric material with two sublattices \(s=A,B\) with locally broken inversion symmetry \cite{FischerAgterbergYanase23, HakunoYanase25}, another possibility is an (interfacial) Dzyaloshinskii-Moriya like interaction (DMI) of the form \(\mathbf{D}_s(\mathbf{q})\cdot\mathbf{M}_s(-\mathbf{q})\times\mathbf{s}_s(\mathbf{q})\) with \(\mathbf{D}_A=-\mathbf{D}_B\) and \(\mathbf{D}_s(\mathbf{q})=-\mathbf{D}_s(-\mathbf{q})\propto\hat{\mathbf{c}}\times\mathbf{q}\) \cite{Valizadeh18, Kuznetsov23}.
In principle, the triplet nonuniform SC order could be characterized either by a nonuniform phase, \(\Delta\sim e^{iqr}\), with fixed Cooper pair spin \(\mathbf{s}\), or by a nonuniform \(\mathbf{s}\) (and thus \(\mathbf{d}\)) rotating in space, for example forming a helical or conical pattern \cite{LiuJain14, ZhangWang25}. The latter could be favored by DMI, which generally favors the formation of magnetic spirals. 
On general symmetry grounds, we anticipate that the magnetic order in the mixed state also acquires a nonuniform character with corresponding SDW momentum \(q_M\), which could be probed by experiments sensitive to magnetic order. We note that a ground state with mixed magnetic and SC order with \(q=0\) but \(q_M\neq0\) has been considered in the vicinity of an SC to FM phase transition, e.g., in Ref.~\cite{BlountVarma79}.

Finally, we comment on possible extrinsic factors. It should be emphasized that some extrinsic factors necessarily do contribute to the observed SDE: at the relevant high magnetic fields, the $J-V$ characteristics clearly show vortex flow behavior above $J_c$, which must therefore be set by vortex depinning. The observed SDE is therefore a form of a VDE. Moreover, extrinsic effects certainly determine the polarity of the diode effect (i.e., the sign of \(\eta\)), which at fixed field polarity is never seen to reverse in the experiment between subsequent sweeps (as one would expect if inversion symmetry were to be broken spontaneously each time). This is in contrast to field-free SDE with spontaneous TRSB, for which the polarity can typically be switched by training with a small magnetic field. The polarity of SDE with spontaneous ISB, on the other hand, is likely selected by either the sample geometry or contact placement, which is not as easily manipulated. Experiments with more complicated contact geometry may potentially find such polarity switching, which we hope to carry out in future studies.

Some known extrinsic effects may be at play in UTe\(_2\). For example, VDE without any finite momentum pairing is known to occur in superconductor/ferromagnetic thin bilayer heterostructures with in-plane magnetization, possibly due to stray fields near the edges of the bilayer \cite{Carapella09, HouMoodera23, GutfreundBuzdin23}.
From a symmetry point of view, this is an illustration of a possible SDE mechanism due to coexistence of (inhomogeneous) magnetic order with superconductivity, which draws some parallels to the SDE we observe in the SC2+PPM region. However, the particular mechanism identified in Refs.~\cite{Carapella09, HouMoodera23, GutfreundBuzdin23} 
is only applicable to thin films and not bulk 3D samples that we studied. It may be relevant, however, if a layered structure (characterized by momentum \(q\)) emerges microscopically in the mixed SC2+PPM state.
A purely 3D VDE, on the other hand, is known to occur due to sample geometry affecting the length of the flux tubes \cite{SwartzHart67, Moll25}. This mechanism may play a role in the SDE we observe if domains with broken inversion symmetry form in the SC2+PPM, providing a source of asymmetric vortex pinning (cf. \cite{ZhangMoll25}). In this case too, however, the formation of a domain structure with ISB and some characteristic length scale \(1/q\) must take place. Whether a nonuniform state in such a scenario can be considered an FFLO state depends to a large degree on that length scale and the microscopic origin of the domains, which we cannot resolve in the present experiment.

We also stress that the presence of extrinsic effects does not by itself imply the absence of intrinsic effects. The source of the asymmetry of the vortex pinning potential can itself be of intrinsic origin. It is known theoretically, as one would expect on symmetry grounds, that in FFLO states the vortex lattices themselves exhibit translation symmetry breaking reflecting that of the underlying superconducting order parameter \cite{HouzetBuzdin02, YangMacDonald04, PutilovBuzdin24}. 
On the other hand, orbital effects responsible for the vortex lattice formation are also known theoretically to lead to \(\mathbf{d}\)-vector rotation \cite{TsutsumiMachida24, MinamideYanase25} and the stabilization of finite-momentum pairing \cite{Liu17, SundareshVayrynen23}.
Calculations in helical SC states additionally show that the shape of individual vortices lacks inversion symmetry \cite{FuchsManfra22, PutilovBuzdin25}.
We note that translation symmetry breaking vortex stripes \cite{wang2025stripes} and asymmetric vortex profiles lacking inversion symmetry \cite{yang2025spectroscopic, sharma2025observation, yin2025yinyangvortexute2011} have been seen on the surface of UTe\(_2\), but at fields far below the metamagnetic transition. 
Vortex imaging in the SC2+PPM region -- while experimentally challenging -- could be very informative in confirming or ruling out an intrinsic origin of the SDE, as would other imaging probes sensitive to translation symmetry breaking.

\section{Discussion and Outlook}\label{sec:discussion}

We reported an SDE in the SC2+PPM coexistence region of the UTe\(_2\) phase diagram above 33~T with a large diode efficiency \(|\eta|\approx33\%\), and the existence of superconductivity inside the PPM phase was established. The absence of SDE outside of this narrow coexistence region of the phase landscape implies that inversion symmetry is only (spontaneously) broken inside the region, which strongly suggests an intrinsic origin of the SDE from a nonuniform SC state. This makes UTe\(_2\) a rare candidate for realizing a triplet FFLO. Directly sensitive imaging experiments are necessary to conclusively establish the existence of such a state; nevertheless, the discovery of SDE in UTe\(_2\) is an important first step towards its possible later identification.

The potential mechanism we have identified for producing SDE from coexistence between magnetic order and superconductivity may be relevant to other materials where FFLO states have been suggested, for example CeCoIn\(_5\) or organic superconductors in which SDE has also been observed recently \cite{Sato25}. 
In general, our results call for revisiting past FFLO candidates in SDE experiments. As in UTe\(_2\), a combination of intrinsic and extrinsic factors are likely to play a role in such systems, but the presence of spontaneous ISB within the SC phase may be helpful in isolating the intrinsic mechanisms. 

Several other systems in which spontaneous symmetry breaking occurs, including ISB, have been found to exhibit SDE. These include Ru-substituted Sr\(_2\)RuO\(_4\) \cite{Hooper04, KaneyasuSigrist10},  the kagome CsV\(_3\)Sb\(_5\) \cite{LeLin24, Lou26, LeLin25}, FeSe \cite{HinlopenMoll26}, and Mo\(_2\)C \cite{Gao26}. In all of these, ISB has been attributed to the formation of asymmetric domain networks, but without a clear explanation of the origin of the asymmetry. It is possible that in some of these systems the ISB takes place due to intrinsic material properties. 
Finally, we mention that given observations of signatures of PDW and FFLO states -- as well as theoretical predictions of SDE \cite{IshizukaYanase21, ChazonoYanase23, ShafferChichinadzeLevchenko24} -- in other parts of the UTe\(_2\) phase diagram, multiple FFLO states of distinct origin may yet be observed in UTe\(_2\) itself.

\begin{acknowledgments}
This project was supported by the EPSRC of the UK through grants EP/X011992/1, EP/Z533695/1, and EP/R513180/1. We acknowledge financial support by the Czech Science Foundation GAČR under the Junior Star Grant No. 26-21795M (STiUS). We acknowledge the support of HFML-RU/NWO-I, member of the European Magnetic Field Laboratory (EMFL) and the Engineering and Physical Sciences Research Council (EPSRC, UK) via its membership to the EMFL (grant EP/N01085X/1). A portion of this work was performed at the National High Magnetic Field Laboratory, which is supported by National Science Foundation Cooperative Agreement No. DMR-2128556 and the State of Florida. The work of D.S. was financially supported by the NSF Quantum
Leap Challenge Institute for Hybrid Quantum Architectures and Networks Grant No. OMA-2016136. D.V.C.
acknowledges financial support 
from Washington University in St. Louis through the Edwin Thompson Jaynes Postdoctoral Fellowship. A.G.E. acknowledges support from Sidney Sussex College (University of Cambridge).

\end{acknowledgments}

\section*{Data availability}
The datasets supporting the findings of this study will be uploaded to the University of Cambridge Apollo Repository prior to publication.

\section*{Competing interests}
The authors declare no competing interests.

\appendix

\section{Experimental methods}\label{app:methods}

\subsection{Sample preparation}

Single crystal UTe$_2$ specimens were grown in a flux of molten salts~\cite{PhysRevMaterials.6.073401MSF_UTe2} by the recipe detailed in Ref.~\cite{Eaton2024}. Samples were contacted by spot-welding four gold wires, each of 50~$\upmu$m diameter, for respective current and voltage leads. These were subsequently connected to copper twisted pairs, which were soldered to the wiring of the probe. The sample investigated in this study had $J$ sourced along the [100] direction, with leads affixed to the (001) surface. Residual resistance ratio (RRR) measurements were performed using a Quantum Design Physical Properties Measurement System in Cambridge with low frequency ac excitations. The RRR value for the sample is $\sim$100. 

\subsection{Electrical transport measurements}
The (non-)reciprocal $J-V$ properties of UTe$_2$ were investigated by performing direct current measurements using a Keithley 6221--2182A combined current source--multimeter system. The in-built pulse delta sweep mode functionality was utilized. Measurements were performed in a 37~T resistive magnet with a $^3$He cryostat system at the EMFL-HFML, Nijmegen, The Netherlands. Control of the instrument was realized by custom-developed Python code with the pymeasure package~\cite{Pymeasure_source}. For each $V(J)$ data point, the current pulse excitation was applied for 5~ms and then turned off for 1~s to reduce Joule heating.  For sweeping current measurements, up and down sweeps were performed to check the overlap. No indicative signatures of Joule heating were observed. The rectification measurement shown in Fig.~\ref{fig:rect} was acquired using the pulse-delta mode feature. The resistivity measurements in swept magnetic field were performed in the same resistive magnet $^3$He cryostat system, in which low-frequency ac measurements were obtained using a Keithley 6221 current source and a Stanford Research 860 series lock-in amplifier. The alignment of the field along the [010] direction was achieved by minimizing the metamagnetic transition field by rotating the sample. For all figures in the main text, the magnetic field was oriented along the [010] direction within 2 degrees.

\begin{figure*}[t]
\centering
\includegraphics[width=\textwidth]{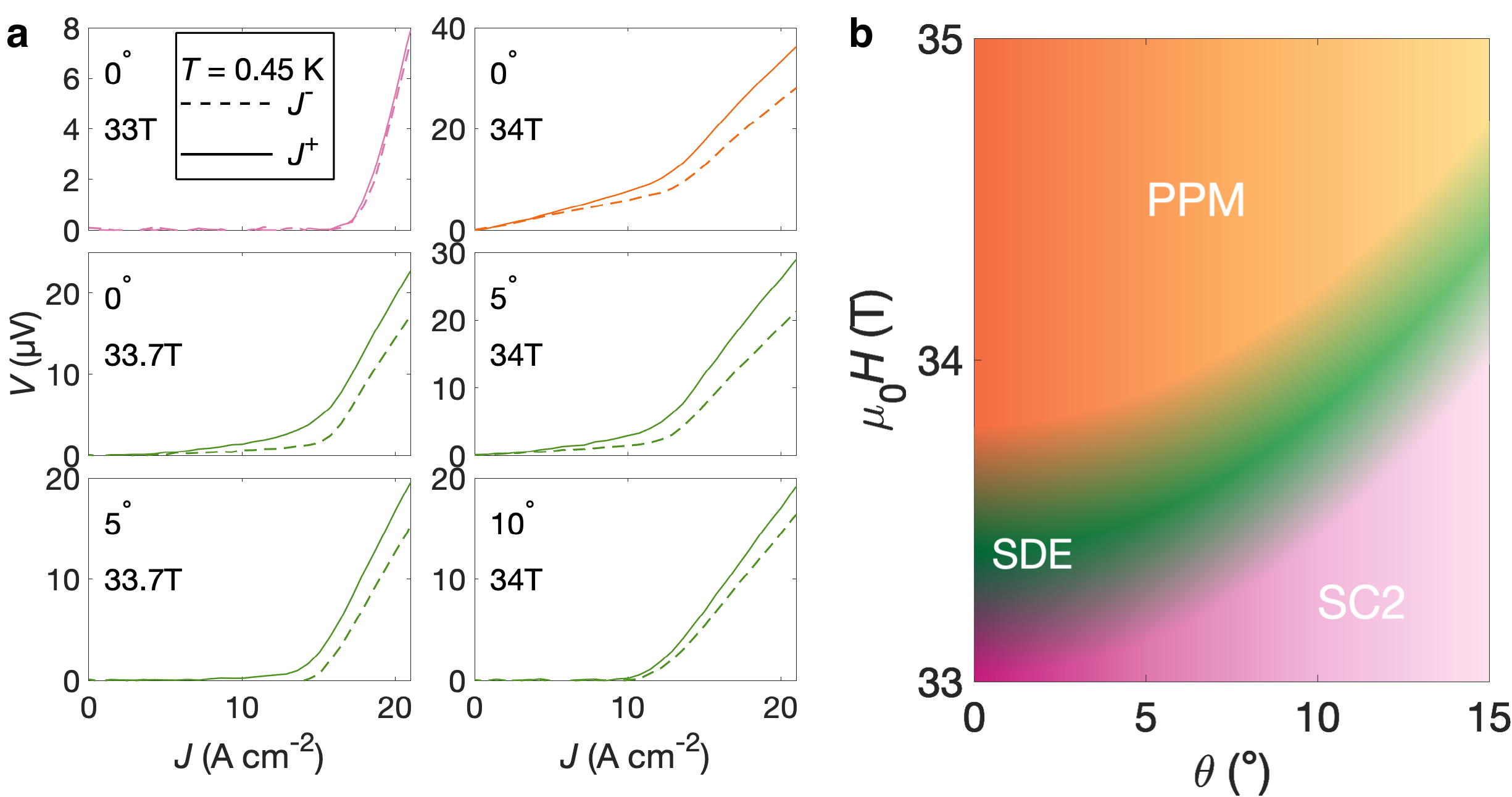}
\caption{\textbf{Tracking the SDE upon tilting the magnetic field.} (a) Positive (solid) and negative (dashed) polarity dc current sweeps at various field strengths and tilt angles $\theta$, from [010] towards [001], as indicated. The nonreciprocity in $V(J)$ tracks the metamagnetic transition field as it migrates to higher $H$~\cite{Ranfieldboostednatphys2019,tony2024enhanced}. (b) Schematic phase diagram of $H$ vs $\theta$ showing how the SDE migrates to higher $H$ with increasing $\theta$ (for small rotations away from [010]).}
\label{Fig:rotate}
\end{figure*}

In Fig.~\ref{Fig:rotate} we present data acquired upon rotating the orientation of the field. The nonreciprocity that characterizes the SDE clearly tracks the metamagnetic transition as it migrates up to higher field strengths upon tilting the field away from [010] towards [001], which has been studied in detail in prior experiments~\cite{Ranfieldboostednatphys2019,lewin2023review,helm2024}.

\begin{figure*}[t]
\centering
\includegraphics[width=0.75\textwidth]{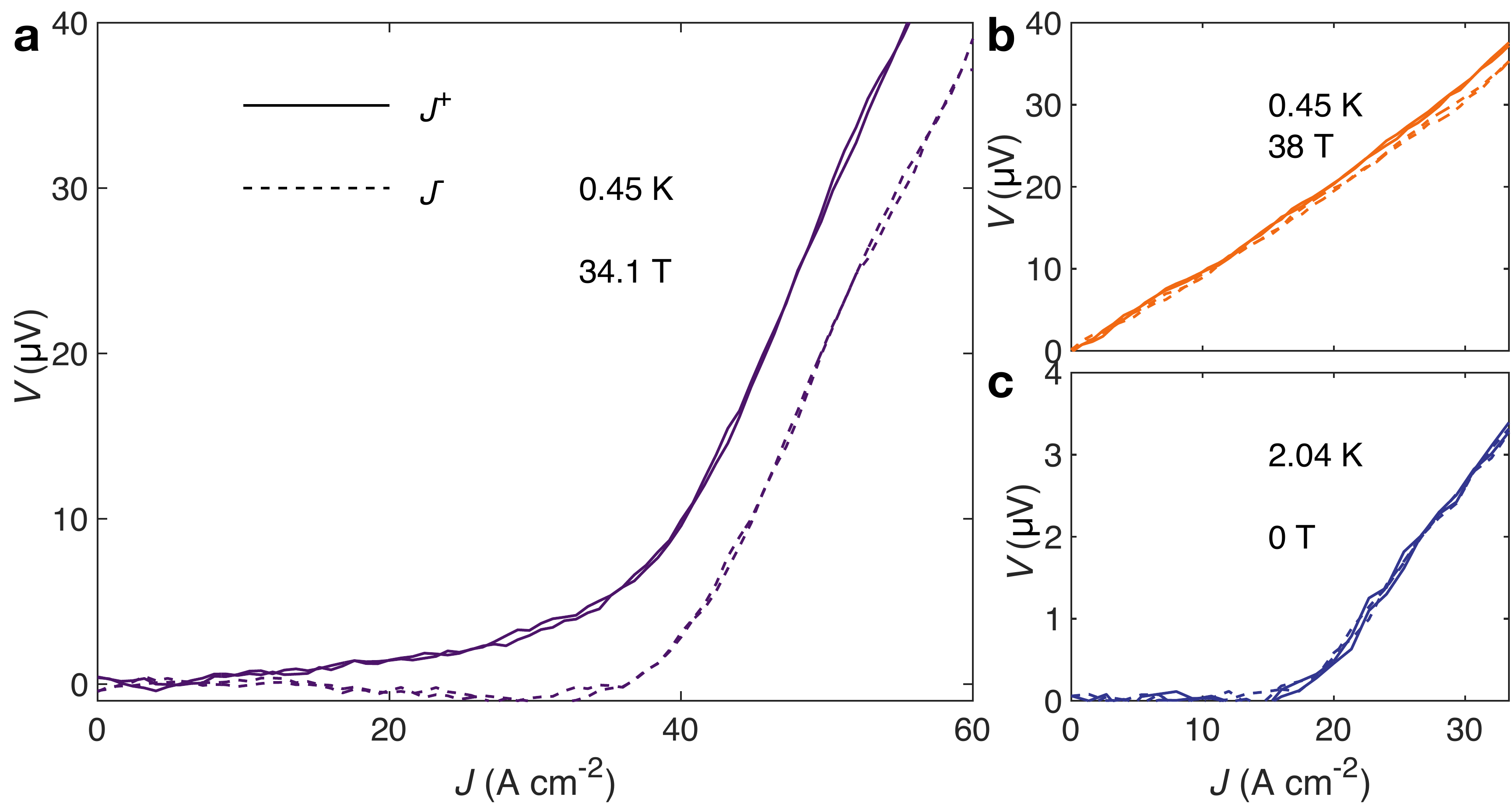}
\caption{\textbf{SDE in sample S2.} Similar voltage-current measurements were performed on a different sample S2.  (a) Superconducting diode effect observed at metamagnetic transition. (b) Normal metal behavior in the polarized paramagnetic state at 38~T. (c) Reciprocal superconducting transition at 0~T. }
\label{Fig:S2}
\end{figure*}

The SDE is also observed at the metamagnetic transition in a second sample, S2 (Fig.~\ref{Fig:S2}), with the magnetic field oriented within 5 degrees of [010]. The measurement was performed in a 41.5~T resistive magnet with a $^3$He cryostat system at the NHMFL, Tallahassee, USA. $V(J)$ at 38~T in the polarized paramagnetic state is Ohmic and reciprocal. Additional $V-J$ measurements at 0~T show a reciprocal superconducting transition. 

\section{GL Model Parameters and SDE computations}\label{app:GL}

The \(|\Delta|^2\) coefficients in the GL free energy in Eq.~\eqref{F} were taken to be \(a_0(T)=a'_0(T-T_c)\) and \(a_H(H)=a_{H0}H^2+a_{fluc}\,\Gamma^2(H-H_m)^2/[(H-H_m)^2+\Gamma^2]^2\,[1-\Theta(H-H_m)]\), where \(\Theta\) is the Heaviside step function and fields are expressed in tesla; \(a_{fluc}\) originates from the fluctuation mediated interactions studied in Ref.~\cite{tony2024enhanced} (and which vanish inside the spin-polarized phase), while \(a_{H0}\) includes the Pauli spin-breaking effect but can also qualitatively account for orbital depairing. 
To obtain Fig.~\ref{Fig:order}, the parameter values were taken to be \(a'_0=15\), \(T_c-T=1\), \(a_{H0}=0.01\), \(a_{fluc}=-40\), \(H_m=33\), \(\Gamma=1\), \(a_2=1\), \(a_4=5\), \(b=0.6\), \(\chi^{-1}=160.616\), \(\beta=-544.008\), \(\delta=634.661\), \(\chi^{-1}_\Delta=444.49\), and \(\gamma_2=-66\).\begin{figure*}[t]
\centering
\includegraphics[width=0.99\textwidth]{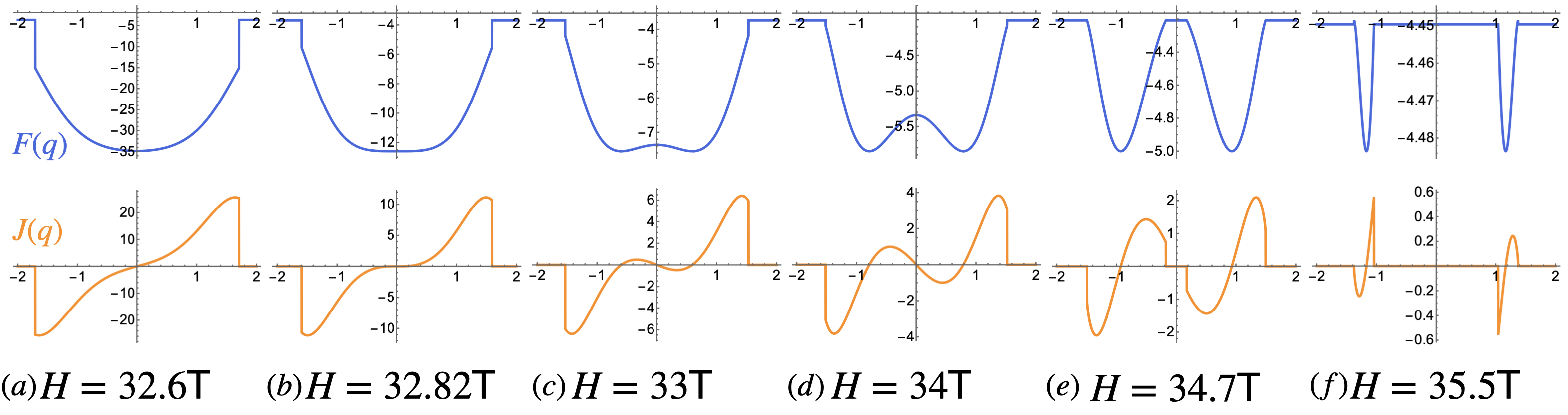}
\caption{\textbf{Condensation energy and supercurrent in the GL model.} Condensation energy \(F(q)\) and the corresponding supercurrent \(J(q)\) corresponding to the order parameters in Fig.~\ref{Fig:order} for several values of \(H\). For \(H<32.82\)~T, the free energy has a global minimum at \(q=0\), and there is no ISB and no SDE. The critical currents are set by the global maximum and minimum of \(J(q)\). At \(H=32.82\)~T, the minimum at \(q=0\) splits into two minima at \(q=\pm q_0\), resulting in spontaneous ISB. The critical currents are now determined by the local minima/maxima of \(J(q)\) nearest to \(q_0\); as a result, upon decreasing \(H\) towards \(32.82\)~T, \(q_0\rightarrow0\) as do the critical currents closest to \(q=0\), formally resulting in \(\eta\rightarrow\pm1\). At higher \(H\), the minimum at \(q=0\) becomes unstable (i.e., there is no SC solution with \(q=0\)), and the range of \(q\) for which SC is possible shrinks until it vanishes at \(H\sim 36\)~T. \(\eta\rightarrow0\) at this transition, as one would expect.}
\label{Fig:current}
\end{figure*}

To compute the SDE, we find the condensation energy \(F(q)=\mathcal{F}_*\), the free energy optimized over \(\Delta\) and \(M\) at fixed \(q\), and compute the supercurrent as \(J(q)=2e\partial_qF\). Generically, one finds that \(F(q)\) has a single global minimum at \(q=0\) at low \(H\), which splits into two at \(q=\pm q_0\) at some \(H^*\) (see Fig.~\ref{Fig:current}). Due to inversion symmetry, the global maximum and minimum of \(J(q)\) are always exactly opposite, but once inversion is broken spontaneously the critical currents \(J_c^\pm\) correspond instead to local extrema of \(J(q)\) nearest to \(q_0\). The diode efficiency coefficient is then defined as \(\eta=(J_c^+ + J_c^-)/(J_c^+ - J_c^-)\), with \(J_c^-<0\) [equivalent to the definition in Sec.~\ref{sec:intro}]. Formally, we find that \(\eta\) approaches its maximum possible value of \(1\) as \(q_0\rightarrow0\) with decreasing \(H\). This is because the two minima merge and \(J_c^-\rightarrow 0\). However, this ignores the fact that the free energy barrier between the two minima also shrinks to zero as \(q_0\rightarrow0\); as a result, as the current is lowered below \(J_c^-\), the system can jump over the barrier instead of entering a resistive state, and remain superconducting until \(J\) reaches its global minimum. In the experiment, one thus expects \(\eta\) to be large but below its perfect value close to the transition into the mixed phase.

\section{Symmetry considerations for the form of the order parameter}\label{app:sym}

Here we discuss in more detail the possible forms of the order parameter presented in Table~\ref{table:data} below, including in particular the relevant irreducible representations (irreps) it may belong to. The presence of multiple irreps is generally required in any scenario to explain the SDE in our model. Throughout Appendixes~\ref{app:sym} and~\ref{app:symq} we use Cartesian axes $(x,y,z)$ along the crystallographic $(a,b,c)$ axes; $\mathbf{q}$ denotes the Cooper-pair (center-of-mass) momentum, which enters the GL free energy, while $\mathbf{k}$ denotes the relative momentum on which the $\mathbf{d}$ vector depends.
To make further comments on the symmetry of \(\Delta\), it is necessary to know the irrep that the magnetization order parameter \(M\) belongs to, 
which is determined by the direction of the magnetization jump at the metamagnetic transition. 
The most likely scenario is that the magnetization is aligned in the same direction as the external field when the latter is aligned along the \(b\) axis (otherwise additional symmetry breaking must take place at the metamagnetic phase transition). This would correspond to \(M\) belonging to the \(B_{2g}\) irrep of \(D_{2h}\). The \(q^2\) in \(\gamma\) can itself originate from different components of \(\mathbf{q}\), e.g., \(q_x q_y\propto q^2\), etc.
Given that \(\Delta_n\Delta_m^*\sim|\Delta|^2\) must contain a non-trivial irrep in order to break TRS, the scenario in which \(q^2\) itself belongs to \(B_{2g}\) (i.e., \(q^2\sim q_x q_z\)) is unlikely.
One possibility is that \(q^2\) belongs to a trivial irrep (i.e., \(q^2\sim q_x^2,q_y^2,q_z^2\)): in that case either \(A_u+iB_{2u}\) or \(B_{1u}+iB_{3u}\) pairing could produce the desired coupling. Alternatively, \(q^2\) may belong to \(B_{1g}\) (\(q^2\sim q_x q_y\)) or \(B_{3g}\) (\(q^2\sim q_y q_z\)). The former would require either \(A_u+iB_{3u}\) or \(B_{1u}+iB_{2u}\) pairing; the latter would require either \(A_u+iB_{1u}\) or \(B_{2u}+iB_{3u}\) pairing.
Depending on the irrep \(q^2\) belongs to, any \(|\Delta|^2\) may contain any non-trivial pair of irreps among \(A_u, B_{1u}, B_{2u}\), or \(B_{3u}\). All the possibilities are summarized in Table~\ref{table:data}.
Nevertheless, note that the irrep of \(q^2\) also determines the directional dependence of the diode effect: if \(q^2\sim q_x q_y\), no diode effect would be expected for a current \(\mathbf{J}\) along the \(c\) axis, etc. Since the diode effect is seen at least for \(\mathbf{J}\) along \(a\), this rules out \(q^2\) belonging to \(B_{3g}\), which in turn rules out \(A_u+iB_{1u}\) and \(B_{2u}+iB_{3u}\) pairing (at least within our model). Measuring the critical currents for \(\mathbf{J}\) along \(b\) and \(c\) axes (and generally measuring the critical current anisotropy) may further constrain the form of the gap function.

\subsection{Details of symmetry analysis without momentum dependence}

In this section we consider possible couplings between the magnetic order and the SC order from the symmetry perspective. UTe$_2$ belongs to the $D_{2h}$ point group. According to the representation table, magnetic order here belongs to $B_g$ irreducible representations (irreps) in the following way:
\begin{eqnarray}
    M_x \in B_{3g},\\
    M_y \in B_{2g},\\
    M_z \in B_{1g}.
\end{eqnarray}
Within the GL approach, symmetry-allowed terms constituting the free energy functional belong to the $A_g$ irrep (i.e., are scalars). 

In UTe$_2$, superconducting (SC) order is widely believed to be spin-triplet. Hence, possible irreps describing SC order parameter include $A_u, B_{1u}, B_{2u}, B_{3u}$. According to the irrep product table, $A_g$ irrep can be constructed out of a product of $B_g$ irreps as follows:
\begin{eqnarray}
    B_{1g} \otimes B_{1g} = A_g, \\
    B_{2g} \otimes B_{2g} = A_g, \\
    B_{3g} \otimes B_{3g} = A_g.
\end{eqnarray}
This indicates, that depending on the magnetization direction in the sample, there will be different relevant SC irreps. Additionally, for a $M \Delta^2$ term to be present in the GL free energy functional, the product of two SC irreps should give a certain $B_g$ irrep. Since magnetization is described by a vector field, we should consider terms of the form
\begin{equation}
    \vec{M} \cdot \left( \vec{d} \times \vec{d}^* \right),
\end{equation}
which justifies the choice of a non-unitary state.

\subsubsection{$M_y$-magnetism}
\label{sym_Sy}

We first consider the case of magnetization developing along the $y-$direction, as this case might be the most experimentally motivated. In such case, the product of two SC irreps should result in $B_{2g}$ irrep. According to the irrep product table, the only two options are
\begin{eqnarray}
    A_{u} \otimes B_{2u} = B_{2g}, \\
    B_{1u} \otimes B_{3u} = B_{2g}. 
\end{eqnarray}

Consider the $A_{u} \otimes B_{2u}$ product first. 
For $A_u$ irreducible representation the superconducting $\vec{d}$-vector is given by 
\begin{equation}
    \vec{d}(\vec{k}) \sim \left( \hat{x} k_x, \hat{y} k_y, \hat{z} k_z \right),
\end{equation}
while for $B_{2u}$ 
\begin{equation}
    \vec{d}(\vec{k}) \sim \left( \hat{x} k_z, \hat{y} k_x k_y k_z, \hat{z} k_x \right).
\end{equation}
Without loss of generality, let us consider a state 
\begin{equation}
    \vec{d}(\vec{k}) = \psi_{A_u} + \alpha \psi_{B_{2u}},
\end{equation}
where $\alpha$ is a complex number and $\psi_{A_u},\psi_{B_{2u}}$ fields describe superconducting condensates belonging to the $A_u$ and $B_{2u}$ irreps respectively. We restrict ourselves to the lowest harmonics in the basis function expansion and calculate
\begin{equation}
    \begin{gathered}
        \vec{d} \times \vec{d}^* = 2 i \text{Im}(\alpha) \left( -k_x k_y, k_x^2 - k_z^2, k_y k_z \right).
    \end{gathered}
\end{equation}

We now proceed to the $B_{1u} \otimes B_{3u}$ product case. 
For $B_{1u}$ the $\vec{d}$-vector reads 
\begin{equation}
    \vec{d}(\vec{k}) \sim \left( \hat{x} k_y, \hat{y} k_x, \hat{z} k_x k_y k_z \right),
\end{equation}
while for $B_{3u}$ 
\begin{equation}
    \vec{d}(\vec{k}) \sim \left( \hat{x} k_x k_y k_z, \hat{y} k_z, \hat{z} k_y \right).
\end{equation}
Without loss of generality, consider a state 
\begin{equation}
    \vec{d}(\vec{k}) = \psi_{B_{1u}} + \alpha \psi_{B_{3u}},
\end{equation}
where $\alpha$ is a complex number. We restrict ourselves to the lowest harmonics in the basis function expansion and calculate
\begin{equation}
    \begin{gathered}
        \vec{d} \times \vec{d}^* = 2 i \text{Im}(\alpha) \left( -k_x k_y, k_y^2, - k_y k_z \right).
    \end{gathered}
\end{equation}


\subsubsection{$M_x$-magnetism}
\label{sym_Sx}

Here and below we consider two other directions of magnetization. Vector field describing magnetization along the $x$-axis ($M_x$) belongs to $B_{3g}$ irrep.
According to the irrep product table, the only two options are
\begin{eqnarray}
    A_{u} \otimes B_{3u} = B_{3g}, \\
    B_{1u} \otimes B_{2u} = B_{3g}. 
\end{eqnarray}

We again consider possible products of superconducting order parameters, starting with $A_{u} \otimes B_{3u}$.
For $A_u$ the $\vec{d}$-vector reads
\begin{equation}
    \vec{d}(\vec{k}) \sim \left( \hat{x} k_x, \hat{y} k_y, \hat{z} k_z \right),
\end{equation}
while for $B_{3u}$ 
\begin{equation}
    \vec{d}(\vec{k}) \sim \left( \hat{x} k_x k_y k_z, \hat{y} k_z, \hat{z} k_y \right).
\end{equation}
Without loss of generality, we consider a state 
\begin{equation}
    \vec{d}(\vec{k}) = \psi_{A_u} + \alpha \psi_{B_{3u}},
\end{equation}
where $\alpha$ is a complex number. We restrict ourselves to the lowest harmonics in the basis function expansion and calculate
\begin{equation}
    \begin{gathered}
        \vec{d} \times \vec{d}^* = 2 i \text{Im}(\alpha) \left( -k_y^2 + k_z^2, k_x k_y, - k_x k_z \right).
    \end{gathered}
\end{equation}

We now proceed to the case of $B_{1u} \otimes B_{2u}$ product.
The $\vec{d}$-vector for $B_{1u}$ irrep reads
\begin{equation}
    \vec{d}(\vec{k}) \sim \left( \hat{x} k_y, \hat{y} k_x, \hat{z} k_x k_y k_z \right),
\end{equation}
while for $B_{2u}$ irrep case we have
\begin{equation}
    \vec{d}(\vec{k}) \sim \left( \hat{x} k_z, \hat{y} k_x k_y k_z, \hat{z} k_x \right).
\end{equation}
Without loss of generality, let us consider a state 
\begin{equation}
    \vec{d}(\vec{k}) = \psi_{B_{1u}} + \alpha \psi_{B_{2u}},
\end{equation}
where $\alpha$ is a complex number. Restricting ourselves to the lowest harmonics in the basis function expansion we obtain
\begin{equation}
    \begin{gathered}
        \vec{d} \times \vec{d}^* = 2 i \text{Im}(\alpha) \left( -k_x^2, k_x k_y, k_x k_z \right).
    \end{gathered}
\end{equation}


\subsubsection{$M_z$-magnetism}
\label{sym_Sz}

Finally, we consider the case of magnetization along the $z$-axis ($M_z$), which belongs to $B_{1g}$ irrep.
According to the irrep product table, the only two options for the superconducting order parameter product are
\begin{eqnarray}
    A_{u} \otimes B_{1u} = B_{1g}, \\
    B_{2u} \otimes B_{3u} = B_{1g}. 
\end{eqnarray}

We start by considering the $A_{u} \otimes B_{1u}$ product.
For order parameter describing the $A_u$ irrep the $\vec{d}-$vector reads
\begin{equation}
    \vec{d}(\vec{k}) \sim \left( \hat{x} k_x, \hat{y} k_y, \hat{z} k_z \right),
\end{equation}
while for $B_{1u}$
\begin{equation}
    \vec{d}(\vec{k}) \sim \left( \hat{x} k_y, \hat{y} k_x, \hat{z} k_x k_y k_z \right).
\end{equation}
Without loss of generality, we consider a state 
\begin{equation}
    \vec{d}(\vec{k}) = \psi_{A_{u}} + \alpha \psi_{B_{1u}},
\end{equation}
where $\alpha$ is a complex number. We restrict ourselves to the lowest harmonics in the basis function expansion and obtain
\begin{equation}
    \begin{gathered}
        \vec{d} \times \vec{d}^* = 2 i \text{Im}(\alpha) \left( k_x k_z, - k_y k_z, -k_x^2 +  k_y^2 \right).
    \end{gathered}
\end{equation}


We now proceed to the $B_{2u} \otimes B_{3u}$ product.
For $B_{2u}$ the SC $\vec{d}-$vector reads 
\begin{equation}
    \vec{d}(\vec{k}) \sim \left( \hat{x} k_z, \hat{y} k_x k_y k_z, \hat{z} k_x \right),
\end{equation}
while for $B_{3u}$ it is given by
\begin{equation}
    \vec{d}(\vec{k}) \sim \left( \hat{x} k_x k_y k_z, \hat{y} k_z, \hat{z} k_y \right).
\end{equation}
Without loss of generality, let us consider a state 
\begin{equation}
    \vec{d}(\vec{k}) = \psi_{B_{2u}} + \alpha \psi_{B_{3u}},
\end{equation}
where $\alpha$ is a complex number. As previously, we restrict ourselves to the lowest harmonics in the basis function expansion and calculate
\begin{equation}
    \begin{gathered}
        \vec{d} \times \vec{d}^* = 2 i \text{Im}(\alpha) \left( k_x k_z, k_y k_z, -k_z^2 \right).
    \end{gathered}
\end{equation}

\section{Symmetry considerations including momentum}\label{app:symq}

To incorporate momentum into the symmetry analysis we need to consider terms of the form $q^2 M \Delta^2$, where $q$ is the finite momentum of the Cooper pair.  Magnetization $\vec{M}$, as previously, is described by a vector field. Therefore, to make a scalar out of such terms we need to have two or three vector fields. That leaves us with three distinct possibilities. 

\subsection{Two vector fields}

\subsubsection{Case 1}

If $q^2$ is a scalar, then $\vec{M}$ and $\Delta^2$ have to be vectors. This case we considered in the previous section. 

\subsubsection{Case 2}

Here we consider $\Delta^2$ to be a scalar, i.e., $|\textbf{d}|^2$, while momentum serves as an additional source of a vector entity. We again consider 3 types of magnetization.

\paragraph{$M_y$ magnetization}

$M_y$ belongs to $B_{2g}$ irrep, therefore, $q^2$ should belong to $B_{2g}$ as well. Since momentum transforms like $B_u$ representation, the only option is 
\begin{equation}
    B_{1u} \otimes B_{3u} = B_{2g},
\end{equation}
or 
\begin{equation}
    q^2 \rightarrow q_z q_x.
\end{equation}

\paragraph{$M_x$ magnetization}

$M_x$ belongs to $B_{3g}$ irrep, therefore, $q^2$ should belong to $B_{3g}$ as well. Since momentum transforms like $B_u$ representation, momentum bilinear should satisfy
\begin{equation}
    B_{1u} \otimes B_{2u} = B_{3g},
\end{equation}
or 
\begin{equation}
    q^2 \rightarrow q_z q_y.
\end{equation}

\paragraph{$M_z$ magnetization}

$M_z$ belongs to $B_{1g}$ irrep, therefore, $q^2$ should belong to $B_{1g}$ as well. Since momentum transforms like $B_u$ representation, the momentum bilinear should correspond to the following product of irreps
\begin{equation}
    B_{2u} \otimes B_{3u} = B_{1g},
\end{equation}
therefore, 
\begin{equation}
    q^2 \rightarrow q_y q_x.
\end{equation}

\subsection{Three vector fields}

To get a scalar out of three vector entities we need to make use of the triple product. This term will, thus, be chiral in essence.

\subsubsection{$M_y$ magnetization}

We now need to make a $B_{2g}$ irreducible representation out of a product of 4 irreps. All of them are \emph{ungerade}. For $q^2$ term there are only 9 possibilities: 3 of them give $A_g$ ($\vec{q} \cdot \vec{q}$), the other 6 are symmetric with respect to switching the order of the two momenta, i.e., $q_x q_y = q_y q_x$. Hence, without loss of generality, it is sufficient to consider just 3 cases. 

The product of two momentum irreps can give only $B_{1g}, B_{2g}$ or $B_{3g}$ irreps.

\paragraph{Momentum product gives $B_{1g}$}

This case requires the following product of irreps:
\begin{equation}
    B_{2u} \otimes B_{3u} = B_{1g}
\end{equation}
or
\begin{equation}
    q^2 \rightarrow q_y q_x.
\end{equation}
The product of $q^2$ and magnetization then becomes
\begin{equation}
    B_{1g} \otimes  B_{2g} = B_{3g},
\end{equation}
which means that $\Delta^2 \in B_{3g}$. We considered that case in Sec.~\ref{sym_Sx}.

\paragraph{Momentum product gives $B_{2g}$}

This case comes from the product
\begin{equation}
    B_{1u} \otimes B_{3u} = B_{2g}
\end{equation}
or
\begin{equation}
    q^2 \rightarrow q_z q_x.
\end{equation}
The product of $q^2$ and magnetization then becomes
\begin{equation}
    B_{2g} \otimes  B_{2g} = A_{g},
\end{equation}
which means that $\Delta^2 \in A_{g}$, or just $|\mathbf{d}|^2$. We considered it before.

\paragraph{Momentum product gives $B_{3g}$}

To obtain $B_{3g}$ irrep one needs a product of $B_{1u}$ and $B_{2u}$:
\begin{equation}
    B_{1u} \otimes B_{2u} = B_{3g}.
\end{equation}
Hence, the momentum bilinear has to satisfy
\begin{equation}
    q^2 \rightarrow q_y q_z.
\end{equation}
The product of $q^2$ and magnetization then becomes
\begin{equation}
    B_{2g} \otimes  B_{3g} = B_{1g},
\end{equation}
which means that $\Delta^2 \in B_{1g}$. We considered that case in Sec.~\ref{sym_Sz}.

\subsubsection{$M_x$ magnetization}

In such case, the product of 4 irreps has to result in a $B_{3g}$ irrep.

\paragraph{Momentum product gives $B_{1g}$}

$B_{1g}$ can be constructed from the following product:
\begin{equation}
    B_{2u} \otimes B_{3u} = B_{1g},
\end{equation}
which leads to the momentum bilinear satisfying
\begin{equation}
    q^2 \rightarrow q_y q_x.
\end{equation}
The product of $q^2$ and magnetization then becomes
\begin{equation}
    B_{1g} \otimes  B_{3g} = B_{2g},
\end{equation}
which leads to $\Delta^2 \in B_{2g}$. We considered that case in Sec.~\ref{sym_Sy}.

\paragraph{Momentum product gives $B_{2g}$}

$B_{2g}$ irrep is obtained from a product of $B_{1u}$ and $B_{3u}$ irreps as follows:
\begin{equation}
    B_{1u} \otimes B_{3u} = B_{2g}.
\end{equation}
The momentum bilinear, therefore, has to satisfy
\begin{equation}
    q^2 \rightarrow q_z q_x.
\end{equation}
The product of $q^2$ and magnetization then becomes
\begin{equation}
    B_{2g} \otimes  B_{3g} = B_{1g},
\end{equation}
which means that $\Delta^2 \in B_{1g}$. We considered that case in Sec.~\ref{sym_Sz}.

\paragraph{Momentum product gives $B_{3g}$}

In such case, the relevant product of irreps is 
\begin{equation}
    B_{1u} \otimes B_{2u} = B_{3g}
\end{equation}
and the momentum bilinear satisfies
\begin{equation}
    q^2 \rightarrow q_y q_z.
\end{equation}
The product of $q^2$ and magnetization then becomes
\begin{equation}
    B_{3g} \otimes  B_{3g} = A_{g},
\end{equation}
which means that $\Delta^2 \in A_{g}$, or just $|\mathbf{d}|^2$. We considered it before.

\subsubsection{$M_z$ magnetization}

Finally, we consider the case of magnetization along the $z-$axis. Here one needs to construct a $B_{1g}$ irrep out of a product of 4 irreps.

\paragraph{Momentum product gives $B_{1g}$}

This can be achieved using 
\begin{equation}
    B_{2u} \otimes B_{3u} = B_{1g},
\end{equation}
so that the momentum bilinear satisfies
\begin{equation}
    q^2 \rightarrow q_y q_x.
\end{equation}
The product of $q^2$ and magnetization then becomes
\begin{equation}
    B_{1g} \otimes  B_{1g} = A_{g},
\end{equation}
which means that $\Delta^2 \in A_{g}$, or just $|\mathbf{d}|^2$. We considered it before.

\paragraph{Momentum product gives $B_{2g}$}

In such case, the irrep product involves $B_{1u}$ and $B_{3u}$ irreps and reads
\begin{equation}
    B_{1u} \otimes B_{3u} = B_{2g}.
\end{equation}
Then the momentum bilinear involves the $q_z$ and $q_x$ components:
\begin{equation}
    q^2 \rightarrow q_z q_x.
\end{equation}
The product of $q^2$ and magnetization then becomes
\begin{equation}
    B_{1g} \otimes  B_{2g} = B_{3g},
\end{equation}
which means that $\Delta^2 \in B_{3g}$. We considered that case in Sec.~\ref{sym_Sx}. 

\paragraph{Momentum product gives $B_{3g}$}

Finally, we consider the last case, which involves 
\begin{equation}
    B_{1u} \otimes B_{2u} = B_{3g},
\end{equation}
or, in terms of momentum bilinear, 
\begin{equation}
    q^2 \rightarrow q_y q_z.
\end{equation}
The product of $q^2$ and magnetization then becomes
\begin{equation}
    B_{1g} \otimes  B_{3g} = B_{2g},
\end{equation}
which means that $\Delta^2 \in B_{2g}$. We considered that case in Sec.~\ref{sym_Sy}.

\subsection{Tabular summary of symmetry analyses}

\begin{table*}[h]
\caption{Summary of symmetry-allowed terms and irreducible representations (irreps) of order parameters from our above analyses.
}
\label{table:data}
\setlength{\tabcolsep}{15pt}
\resizebox{\textwidth}{!}{
\begin{tabular}{||c|c|c|c||} 
\hline
\makecell{Cooper pair \\ momentum irrep} & \makecell{Magnetic order \\ parameter irrep} & \makecell{Superconducting order \\  parameter irrep} & \makecell{Superconducting order \\ parameter product} \\ [0.5ex] 
\hline
 & $M_x \in B_{3g}$ & $\begin{cases} \vec{d} = \psi_{A_u} + \alpha \psi_{B_{3u}}  \\  
\vec{d} = \psi_{B_{1u}} + \alpha \psi_{B_{2u}} \end{cases}$ & $\begin{cases} \vec{d} \times \vec{d}^* = 2 i \text{Im}(\alpha) \left( -k_y^2 + k_z^2, k_x k_y, - k_x k_z \right)  \\ \vec{d} \times \vec{d}^* = 2 i \text{Im}(\alpha) \left( -k_x^2, k_x k_y, k_x k_z \right)  \end{cases}$ \\
\cline{2-4}
$q^2 \in A_g$ & $M_y \in B_{2g}$ & $\begin{cases} \vec{d} = \psi_{A_u} + \alpha \psi_{B_{2u}}  \\  
\vec{d} = \psi_{B_{1u}} + \alpha \psi_{B_{3u}} \end{cases}$ & $\begin{cases} \vec{d} \times \vec{d}^* = 2 i \text{Im}(\alpha) \left( -k_x k_y, k_x^2 - k_z^2, k_y k_z \right)  \\ \vec{d} \times \vec{d}^* = 2 i \text{Im}(\alpha) \left( -k_x k_y, k_y^2, - k_y k_z \right)  \end{cases}$ \\
\cline{2-4}
 & $M_z \in B_{1g}$ & $\begin{cases} \vec{d} = \psi_{A_u} + \alpha \psi_{B_{1u}}  \\  
\vec{d} = \psi_{B_{2u}} + \alpha \psi_{B_{3u}} \end{cases}$ & $\begin{cases} \vec{d} \times \vec{d}^* = 2 i \text{Im}(\alpha) \left( k_x k_z, - k_y k_z, -k_x^2 +  k_y^2 \right)  \\ \vec{d} \times \vec{d}^* = 2 i \text{Im}(\alpha) \left( k_x k_z, k_y k_z, -k_z^2 \right)  \end{cases}$ \\
\hline
\multirow{4}{*}{$q_y q_x \in B_{2u} \otimes B_{3u}$} & $M_y \in B_{2g}$ & $\begin{cases} \vec{d} = \psi_{A_u} + \alpha \psi_{B_{3u}}  \\  
\vec{d} = \psi_{B_{1u}} + \alpha \psi_{B_{2u}} \end{cases}$ &  $\begin{cases} \vec{d} \times \vec{d}^* = 2 i \text{Im}(\alpha) \left( -k_y^2 + k_z^2, k_x k_y, - k_x k_z \right)  \\ \vec{d} \times \vec{d}^* = 2 i \text{Im}(\alpha) \left( -k_x^2, k_x k_y, k_x k_z \right)  \end{cases}$ \\
\cline{2-4} 
 & $M_x \in B_{3g}$ & $\begin{cases} \vec{d} = \psi_{A_u} + \alpha \psi_{B_{2u}}  \\  
\vec{d} = \psi_{B_{1u}} + \alpha \psi_{B_{3u}} \end{cases}$ & $\begin{cases} \vec{d} \times \vec{d}^* = 2 i \text{Im}(\alpha) \left( -k_x k_y, k_x^2 - k_z^2, k_y k_z \right)  \\ \vec{d} \times \vec{d}^* = 2 i \text{Im}(\alpha) \left( -k_x k_y, k_y^2, - k_y k_z \right)  \end{cases}$ \\
\hline
\multirow{4}{*}{$q_z q_y \in B_{1u} \otimes B_{2u}$} & $M_z \in B_{1g}$ & $\begin{cases} \vec{d} = \psi_{A_u} + \alpha \psi_{B_{2u}}  \\  
\vec{d} = \psi_{B_{1u}} + \alpha \psi_{B_{3u}} \end{cases}$ & $\begin{cases} \vec{d} \times \vec{d}^* = 2 i \text{Im}(\alpha) \left( -k_x k_y, k_x^2 - k_z^2, k_y k_z \right)  \\ \vec{d} \times \vec{d}^* = 2 i \text{Im}(\alpha) \left( -k_x k_y, k_y^2, - k_y k_z \right)  \end{cases}$ \\
\cline{2-4} 
 & $M_y \in B_{2g}$ & $\begin{cases} \vec{d} = \psi_{A_u} + \alpha \psi_{B_{1u}}  \\  
\vec{d} = \psi_{B_{2u}} + \alpha \psi_{B_{3u}} \end{cases}$ & $\begin{cases} \vec{d} \times \vec{d}^* = 2 i \text{Im}(\alpha) \left( k_x k_z, - k_y k_z, -k_x^2 +  k_y^2 \right)  \\ \vec{d} \times \vec{d}^* = 2 i \text{Im}(\alpha) \left( k_x k_z, k_y k_z, -k_z^2 \right)  \end{cases}$ \\
\hline
\multirow{4}{*}{$q_z q_x \in B_{1u} \otimes B_{3u}$} & $M_x \in B_{3g}$ & $\begin{cases} \vec{d} = \psi_{A_u} + \alpha \psi_{B_{1u}}  \\  
\vec{d} = \psi_{B_{2u}} + \alpha \psi_{B_{3u}} \end{cases}$ & $\begin{cases} \vec{d} \times \vec{d}^* = 2 i \text{Im}(\alpha) \left( k_x k_z, - k_y k_z, -k_x^2 +  k_y^2 \right)  \\ \vec{d} \times \vec{d}^* = 2 i \text{Im}(\alpha) \left( k_x k_z, k_y k_z, -k_z^2 \right)  \end{cases}$ \\
\cline{2-4} 
& $M_z \in B_{1g}$ & $\begin{cases} \vec{d} = \psi_{A_u} + \alpha \psi_{B_{3u}}  \\  
\vec{d} = \psi_{B_{1u}} + \alpha \psi_{B_{2u}} \end{cases}$ & $\begin{cases} \vec{d} \times \vec{d}^* = 2 i \text{Im}(\alpha) \left( -k_y^2 + k_z^2, k_x k_y, - k_x k_z \right)  \\ \vec{d} \times \vec{d}^* = 2 i \text{Im}(\alpha) \left( -k_x^2, k_x k_y, k_x k_z \right)  \end{cases}$ \\
\hline
$q_z q_y \in B_{1u} \otimes B_{2u}$ & $M_x \in B_{3g}$ & \multirow{3}{*}{Any $\vec{d}$}  &  \\
\cline{1-2} 
$q_z q_x \in B_{1u} \otimes B_{3u}$ & $M_y \in B_{2g}$ & & $ \vec{d} \cdot \vec{d}^* $ \\
\cline{1-2} 
$q_y q_x \in B_{2u} \otimes B_{3u}$ & $M_z \in B_{1g}$ &  & \\
\hline
\end{tabular}
}
\end{table*}

\clearpage

\bibliography{UTe2}

\end{document}